\documentclass[preprint]{aastex631}

\begin{document}
\title{Why haven't the observed high-frequency QPOs been produced by (MHD) computer simulations of black hole accretion disks?}
\author{Robert V. Wagoner\altaffiliation{1}  and Celia R. Tandon}
\affil{Dept.\ of Physics and KIPAC, Stanford University, USA}
\altaffiliation{wagoner@stanford.edu}

\begin{abstract}
We compare some predictions of \citet{wt21} (WT) with the results of the hydrodynamic and magnetohydrodynamic (MHD) simulations of \citet{rm09} (RM). It appears that the MHD simulations were not run for long enough and the numerical damping was not small enough to produce the observed high-frequency QPOs (and the g-mode seen in the hydro simulations).  
\end{abstract}

\section{Introduction}

We do not consider the low-frequency QPOs, since their frequencies are not stable. As an example of a possible high-frequency QPO (HFQPO) in a black hole accretion disk, we choose the fundamental (axisymmetric) g-mode seen in the hydro simulations of RM. We shall employ the dimensionless time (g-mode phase) $\tau = 2\pi f_0 t \approx  (2.8\times10^{-2} c^3 /GM)t $, where $f_0 \approx 4.5\times10^{-3} c^3 /GM $ is the frequency of the g-mode. The run time of the canonical MHD simulation MHD{\_}1 of RM was $\tau(\mathrm{run}) = 1.1 \times 10^3 $. 

    The quantity plotted in Fig. 2 of RM is of order the turbulent stresses (typically greater than the magnetic stresses within the eigenfunction of the mode). We have assumed that the divergence of the turbulent stress tensor is the main driver of the mode. From that Figure, it appears that the damping time of the hydro runs was $\tau(\mathrm{damp}) \approx 11$. We (and RM) assume that this is due to numerical dissipation, in which case $\tau(\mathrm{damp}) \approx (\mbox{grid size})^{-1/2}$. (WT have shown that ‘turbulent viscosity’ does not operate on modes of oscillation.) The damping time of the MHD run was about 10 times longer. This is consistent with the fact that the grid sizes of runs HD3d{\_}1 and MHD{\_}1 were similar. 

    \citet{gk} showed that convective turbulence excites (drives) the p-modes in the sun. However, the origin of the damping of these modes is still unknown (references in WT). The possibilities quoted do not include turbulent viscosity.

\section{Results}

    We take the damping rate parameter $D_*$  introduced by WT to be equal to $1/\tau(\mathrm{damp})$. The amplitude of the mode should grow until a time of order 1/(damping rate), as WT found. WT also found that if equilibrium (roughly constant energy $E$ of the mildly nonlinear oscillations) is reached, the short time average is 
\begin{equation}  
    \langle E \rangle \approx \langle A^2 \rangle \sim 0.1(\eta\mathcal{M}^2)^2 / D_*  .         
\end{equation}                                                                                                          
 The (dimensionless) oscillation amplitude $A(\tau)$ is approximately equal to the fluid displacement divided by the size of the smallest dimension of the eigenfunction (of order the thickness $h$ of the accretion disk). The parameter $\eta$ is the volume of the maximally coupled edd(ies) divided by the volume of the eigenfunction, and $\mathcal{M}$ is the Mach number of those eddies. Like the MHD (but not the hydro) runs of RM, our evolutions were begun with $A = dA/d\tau= 0$. Since equilibrium was achieved by RM $[\tau(\mathrm{run}) \gg \tau(\mathrm{damp})]$, we find from equation (1) [with $D_* = 1/\tau(\mathrm{damp}) \sim 0.01$ for the MHD run] that for representative values $\eta\sim\mathcal{M} \sim 0.1$, $\langle A^2 \rangle \sim 10^{-5} $. This amplitude would presumably be undetectable in the MHD simulations. 
    
Employing the observed PSDs of black-hole binaries (BHBs) and narrow-line Seyfert 1 AGNs (NLS1s), WT found that the visibility of a mode requires that 
 \begin{equation} 
\langle A^2 \rangle \gtrsim [10 \mbox{(BHBs)} - 10^2 \mbox{(NLS1s)}] /Q  \: ,    
\end{equation}
where $1/Q$ is the fractional width of the QPO. Most observed HFQPOs have $Q \lesssim 4$. Also, WT verified that nonlinearity reduced the value of $Q$.  We have assumed an axisymmetric mode of extent $\delta r \sim \delta z \sim h(r) \sim 10^{-2} r $. We note that leakage of a mode into a traveling wave \citep{ORM} can reduce the required amplitude $A$.

WT found that a ‘blowout’ (onset of the effects of the lowest order nonlinearity of the restoring force, terminating the run) occurred if the amplitude reached a value $|A| = 1$ (at $E = 1/6$), which from equation (1) occurred if $D_* \lesssim 0.1(\eta\mathcal{M}^2)^2$. Therefore, in order to reach a (more observable) blowout during a run, 
 \begin{equation}
      \tau(\mathrm{damp}) = 1/D_* \gtrsim 10\langle A^2\rangle(\eta\mathcal{M}^2 )^{-2} \sim 10^7 .            
\end{equation}                                    
Therefore, much longer run times [$\tau(\mathrm{run}) \gtrsim \tau(\mathrm{damp})$] are required. 

\section{Conclusions}

    From these results, it would appear that in order for simulations of black hole accretion disks to produce the observed high-frequency QPOs, much smaller grid sizes and consequently much longer runs must be employed. This result applies to QPOs produced by normal modes of oscillation whose minimum extent is of order the local thickness of the accretion disk. Topics for future research then include: (1) How can this finite grid size source of damping be reduced? (2) Why is our predicted damping within black hole accretion disks with observed HFQPOs so small, and what is its origin?
 
\begin{acknowledgments}
     We thank Marek Abramowicz and Cole Miller for comments. CT acknowledges financial support from the Stanford Physics Department Summer Undergraduate Research Program.  We would like to thank Stanford University and the Stanford Research Computing Center for providing computational resources on the Sherlock cluster and support. 
\end{acknowledgments}

\end{document}